\documentclass[twocolumn,amsmath,amssymb, reprint]{revtex4-2}

\usepackage{graphicx}
\usepackage{dcolumn}
\usepackage{bm}
\usepackage{braket}
\usepackage{color}

\usepackage[utf8]{inputenc}
\usepackage[T1]{fontenc}
\usepackage{mathptmx}
\usepackage{etoolbox}

\begin{document}


\title{Second spectrum of charge carrier density fluctuations in graphene due to trapping/detrapping processes}
\author{Francesco M. D. Pellegrino}
\author{Giuseppe Falci}%
\author{Elisabetta Paladino}%
\affiliation{ 
Dipartimento di Fisica e Astronomia "Ettore Majorana", Universit\`{a} di Catania, Via S.~Sofia 64, I-95123 Catania, Italy.\\
INFN, Sez. Catania, I-95123 Catania, Italy. \\
 CNR-IMM, Via S. Sofia 64, I-95123 Catania, Italy.
}%


\begin{abstract}
We investigate the second spectrum of  charge carrier density fluctuations in graphene within the McWorther model, where  noise is induced by electron traps in the substrate. Within this simple picture, we obtain a closed-form expression  including both  Gaussian and non-Gaussian fluctuations. We show that a very extended distribution of switching rates of the electron traps in the substrate leads to a carrier density power spectrum with a non-trivial structure on the scale of the measurement bandwidth. This explains the appearance of a $1/f$ component in the Gaussian part of the second spectrum, which adds up to the expected frequency-independent term. Finally, we find that the non-Gaussian part of the second spectrum can become quantitatively relevant by approaching extremely low temperatures.
\end{abstract}

\maketitle

Statistical properties of fluctuations in solids are a signature of the underlying microscopic processes.
The same fluctuations are responsible for the ultimate accuracy of the measurement of any physical quantity, thus they play a central role in applications. 
For zero mean processes, the lowest order statistics is expressed by  two-times correlation functions and their Fourier transforms, related to the noise spectral density via the Wiener-Khintchine theorem~\cite{wiener_acta_1930,khintchine_1934}. Gaussian random processes are entirely characterized by the noise spectral density. 

Noise detection is routinely performed to characterize the efficiency of solid-state nanodevices. Fluctuations (both classical and quantum) represent the main limiting factor for quantum technological applications based on the fragile properties of quantum coherence and entanglement. Conversely, quantum devices can be employed as highly efficient noise spectrometers, also sensitive to higher-order statistics of fluctuating observables. For instance, superconducting quantum bits whose coherent properties are limited by microscopic material-inherent fluctuations with low-frequency $1/f$ spectral density, also display evidence of superposed Lorentzian spectral lines possibly due to bistable fluctuators of different nature, intrinsically non-Gaussian~\cite{PaladinoRMP,Muller2019}. 
Non-Gaussian contributions in $1/f$ spectral densities are conveniently identified by the second spectrum \cite{Restle1985,weissman_rmp_1988,seidler_prb_1996,kogan_book}, i.e. the spectral density of the squared signal transmitted through a narrow bandpass filter, a quantity related to the four-times correlation function of the fluctuating observable.  Measurements of the second spectrum have been employed to identify statistical properties of $1/f$ noise in different systems~\cite{kogan_book}. 

Noise with $1/f$ spectral density is a ubiquitous phenomenon, reported in graphene for the first time in 2008~\cite{Chen2007,Lin2008} and since then studied in a number of device configurations, substrate types, and for a large range of mobility values~\cite{balandin_natnano_2013,karnatak_advX_2017}. 
Understanding $1/f$ noise in graphene is important both from a fundamental point of view and for the variety of current and envisaged applications based on its remarkable electrical, thermal, and mechanical properties. 
The origin of $1/f$ noise in graphene is complex \cite{balandin_natnano_2013,karnatak_advX_2017}, in particular near the charge degeneracy point where it is argued to arise from different mechanisms, namely charge traps, defects, long and short-range scatterers or charge puddles. Correlations between charge traps and mobility fluctuations have been also found \cite{Pellegrini2013,lu_prb_2014}.
Thanks to encapsulation in hexagonal Boron Nitride (hBN) or suspended configurations, intrinsic fluctuations of pristine graphene have been investigated.
Recent studies of $1/f$ noise in graphene in a magnetic field provided evidence of a mobility fluctuations mechanism~\cite{Kamada2021,rehman_nanoscale_2022,kamada_nanol_2021}.
Furthermore, a study conducted on graphene sandwiched between the hBN substrate in a dual-gated configuration has recently shown that in the electron-doped region of the material, a distinct peak in the noise amplitude is evident at the Fermi level $\mu_0\sim 90$~meV above the charge neutrality point~\cite{kumar_apl_2021}. This peak has been  attributed to impurity trap states arising from carbon replacing nitrogen in the hBN crystal
and it deviates from the usual weak gate-voltage dependence of the longitudinal resistance noise indicating a mobility fluctuations origin~\cite{rehman_nanoscale_2022}.
This finding agrees with an explanation based on the McWhorter charge number fluctuation model~\cite{pellegrino_commphys_2020}, and $1/f$ noise measurements have been proposed as an extremely sensitive tool to probe the defect states in the hBN substrate~\cite{kumar_apl_2021}.
More generally, it is accepted that the external electrostatic fluctuations arising from trap states  take place in basic graphene devices consisting of single-layer graphene exfoliated on a few-hundreds nanometers of oxide grown on top of highly doped silicon \cite{karnatak_advX_2017}. Under these conditions, the proximity of the transport channel to charge traps and reduced screening properties of single-layer graphene, lead to carrier density fluctuations with $1/f$ spectral density~\cite{pellegrino_commphys_2020,pellegrino_jstat_2019}. 
Fluctuations in charge carrier density in a two-dimensional electron gas
reflects in fluctuations of directly related quantities, 
such as the critical current in a short ballistic graphene Josephson junction \cite{pellegrino_commphys_2020} and Hall voltage in the presence of a non-quantizing, static, and transverse low magnetic field~\cite{lu_prb_2014,brophy_pr_1957,kurdak_prb_1997}. In both examples, mobility does not play an active role.
Finally, control of the carrier density noise~\cite{lu_prb_2014} can be useful to conceive possible strategies for suppressing the phenomena of relaxation and decoherence in qubits based on ballistic graphene Josephson junctions~\cite{pellegrino_commphys_2020,pellegrino_proc_2019,Casparis_NatNanotech_2018, wang_natnanotech_2019}.

Motivated by these considerations, in this work we evaluate the second spectrum of charge carrier density fluctuations in graphene
due to charge traps described by the McWorther model. 
\begin{figure}[t]
\centering
\includegraphics[width=\columnwidth]{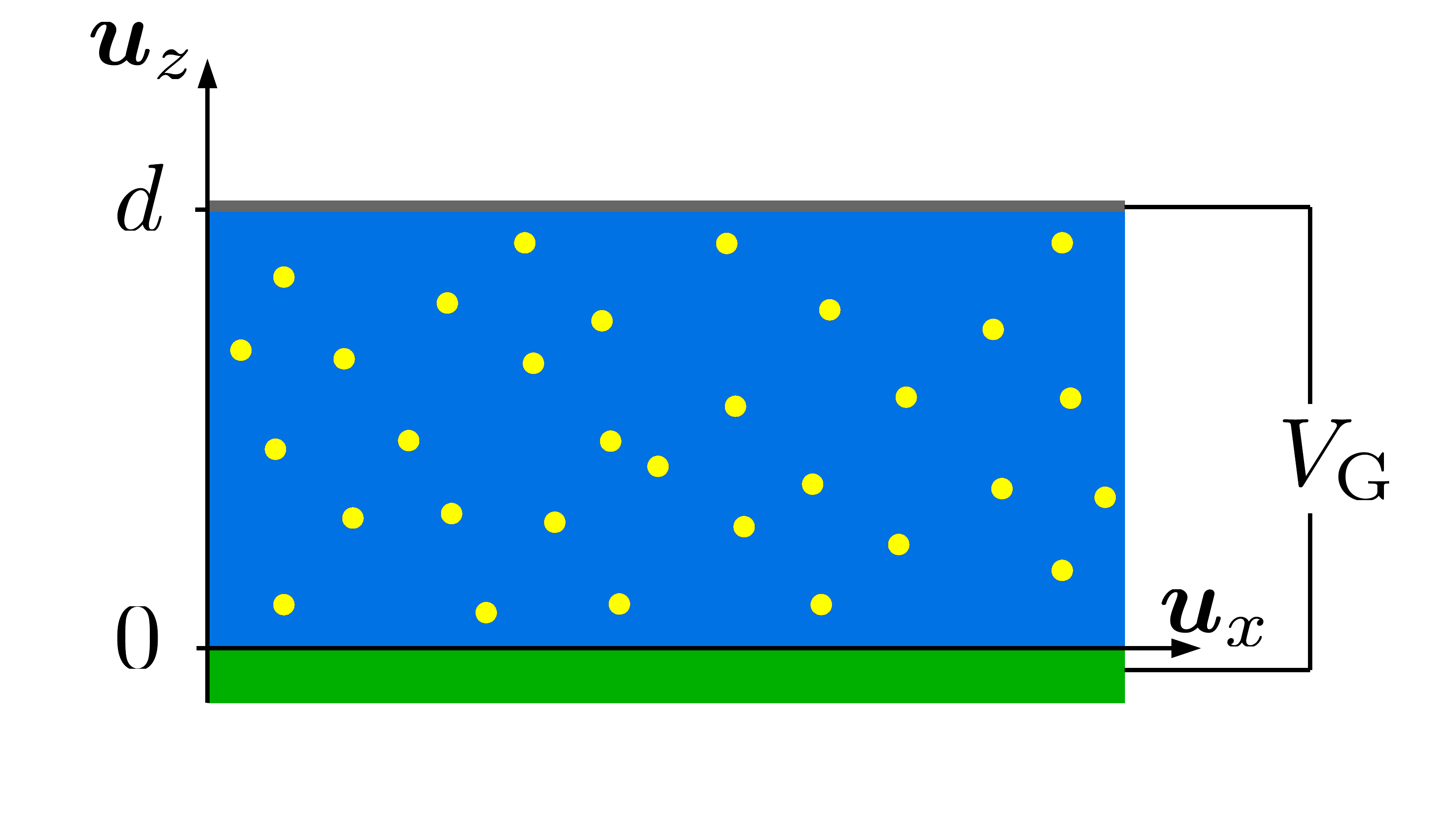}
\caption{Schematics of the device (side view). The device consists of
a metal gate (green), a substrate (blue), and a monolayer graphene (gray). Within the substrate, the yellow circles denote the electron traps. Between the monolayer graphene and the metallic gate is forced a voltage drop $V_G$ to set the average charge carrier density in graphene.
}
\label{fig:setup}
\end{figure}
We consider the simple device shown in Fig.~\ref{fig:setup}, from  top to bottom, it is formed by a single layer of graphene placed at $z=d$, a substrate that contains defects which act as electron traps, and a metal gate.
In this setup, the metal-substrate interface is at $z=0$, 
and the width of the substrate $d$ is much smaller than the lengths of the $\bm{u}_x$ and $\bm{u}_y$ directions ($L_x$ and $L_y$). The voltage drop between the metal gate and the graphene layer $V_{\rm G}$ is fixed as
\begin{equation}
 V_{\rm G}=V_{\rm T}+V_{\rm g}+\frac{W_{\rm f}}{e}~,   
\end{equation}
where $V_{\rm T}$ and  $V_{\rm g}$  are the voltage drops due to trapped charges in the substrate and due
to carrier density in graphene, respectively, and $W_{\rm f}$ is the work function difference between the gate and graphene.
In the large doping regime,  the electro-chemical potential $V_{\rm g}=e n/C_{\rm g}$ consists only of the electrostatic contribution~\cite{fernandez_prb_2007}, where $n$ is the charge carrier density in graphene, and $C_{\rm g}=\varepsilon_{\rm r}/(4 \pi d)$ is the geometric capacitance in terms of the relative dielectric constant.
Here, charge variations in the substrate traps occur due to trapping-recombination events, which are treated as separate Markov processes, and the time between two consecutive transitions is considered much greater than the time it takes the crystal to reach equilibrium after each transition~\cite{kogan_book}.
Within this system, charge density in graphene responds instantaneously to the ﬂuctuations of the density of populated trap states per unit volume and energy $\Delta{\cal N}_{\rm T}(\epsilon,{\bm r},z,t)$ around its average value ${\cal N}_{\rm T0}(\epsilon,{\bm r},z)$. By keeping fixed the gate voltage $V_{\rm G}$, one has
\begin{equation}\label{eq:dn}
 \Delta n(t)=\frac{C_{\rm g}}{-e} \Delta V_{\rm T}=-
 \int_0^d d z  \frac{z}{d} \int \frac{d {\bm r}}{L_x L_y}\int^\Lambda_{-\Lambda} d \epsilon\Delta {\cal N}_{\rm T}(\epsilon,{\bm r},z,t)~,
\end{equation}
where $({\bm r},z)$ is the position vector, $\Lambda$ is the cut-oﬀ energy. We suppose that carrier traps are homogeneously distributed  both in the $\bm{u}_x$ and $\bm{u}_y$ directions~\cite{pellegrino_jstat_2019}, and 
\begin{equation}\label{eq:Nt}
\Delta {\cal N}_{\rm T}(\epsilon,{\bm r},z,t)=\sum^{M_{\rm T}}_{i=1}\delta(\epsilon-\epsilon_i) \delta({\bm r}-{\bm r}_i)     \delta(z-z_i) x(i,t), 
\end{equation}
where $M_{\rm T}$ is the total number of traps, $x(i,t)=X(i,t)-f_i$, and $X(i,t)$ is a telegraph process~\cite{kogan_book} randomly switching with the time-independent rate $\gamma_i$ (stationary process) being one (zero) when trap $i$ is filled (empty). 
The stationary probability that the trap $i$ is occupied is $f_i$. 
We consider that trap $i$ is located at position $(\bm{r}_i, z_i)$ and the energy of the occupied trap is $\epsilon_i$ (by setting the zero energy at the charge neutrality point of graphene).

Within this description based on the McWhorter model, we calculate the second spectrum~\cite{kogan_book,schad_prb_2014} of the carrier density fluctuations in graphene, $\Delta n(t)=n(t)-\braket{n}$, which is defined as
\begin{equation}
\label{eq:S2}
\begin{aligned}
&S_{n}^{(2)}(\omega|\omega_0,\Delta \omega)= \lim_{T_t \to \infty}
\frac{8}{T_t} 
\int^{T_t/2}_{-T_t/2} d \tau_1 e^{i \omega \tau_1} 
\int^{T_t/2}_{-T_t/2} d \tau_2 e^{-i \omega \tau_2}\\
&\times  \braket{ \Delta S_n(\tau_1|\omega_0,\Delta \omega) \Delta S_n(\tau_2|\omega_0,\Delta \omega)},
\end{aligned}
\end{equation}
where the total measurement time $T_t$ is the longest timescale.
The carrier density noise signal measured at time $\tau$  by a spectrum analyzer with filter frequency $\omega_0$ and  bandwidth $\Delta \omega=2\pi/T_{\rm s}$ determined by the duration $T_s$ of a single measurement, is expressed as follows 
\begin{equation}
\label{eq:dS}
\begin{aligned}
\Delta S_n(\tau|\omega_0,\Delta \omega)&= \frac{1}{T_s^2}   
\int^{\tau+T_s/2}_{\tau-T_s/2}
d t_1 e^{i \omega_0  t_1}
\int^{\tau+T_s/2}_{\tau-T_s/2}  e^{-i \omega_0  t_2}\\
&\times[\Delta n (t_1) \Delta n(t_2)-\braket{\Delta n (t_1) \Delta n(t_2)}]~.
\end{aligned}
\end{equation}
By using Eqs.~\eqref{eq:dn} and \eqref{eq:Nt}, we can write the multi-time correlators of the carrier of density fluctuations $\Delta n(t)$. We will calculate the fourth-order correlators for obtaining the expression of $S^{(2)}_n(\omega|\omega_0,\Delta \omega)$.

The second-order correlator of the carrier of density fluctuations~\cite{pellegrino_jstat_2019,pellegrino_commphys_2020} is expressed as
\begin{equation}\label{eq:n1_n2}
\begin{aligned}
&\braket{\Delta n(t_1)\Delta n(t_2)}  =  \frac{1}{L_x^2 L_y^2} 
\sum^{M_{\rm T}}_{i=1} \frac{z_i^2}{d^2}f_{i}(1-f_{i})e^{-\gamma_{i}|t_1-t_2|}.
\end{aligned}
\end{equation}
Exploiting Markovianity~\cite{kogan_book}, and assuming that traps are uncorrelated, we can calculate any multi-time correlators in terms of $f_i$ and $\gamma_i$ (details are in the supplemental material).
For the sake of clarity,  we split the fourth-order multi-time correlator into the Gaussian and the non-Gaussian part as follows:
\begin{equation}\label{eq:n1_n4}
\begin{aligned}
\braket{\Delta n(t_1)\Delta n(t_2)\Delta n(t_3)\Delta n(t_4)}&=\braket{\Delta n(t_1)\Delta n(t_2)\Delta n(t_3)\Delta n(t_4)}_{\rm G}\\
&+\braket{\Delta n(t_1)\Delta n(t_2)\Delta n(t_3)\Delta n(t_4)}_{\rm NG}.
\end{aligned}
\end{equation}
The Gaussian part is defined as
\begin{equation}\label{eq:n1_n4G}
\begin{aligned}
&\braket{\Delta n(t_1)\Delta n(t_2)\Delta n(t_3)\Delta n(t_4)}_{\rm G}=\braket{\Delta n(t_1)\Delta n(t_2)}\braket{\Delta n(t_3)\Delta n(t_4)}\\
&+\braket{\Delta n(t_1)\Delta n(t_3)}\braket{\Delta n(t_2)\Delta n(t_4)}\\
&+\braket{\Delta n(t_1)\Delta n(t_4)}\braket{\Delta n(t_2)\Delta n(t_3)}~,
\end{aligned}
\end{equation}
and it is completely expressed in terms of the second-order correlator shown in Eq.~\eqref{eq:n1_n2}.
The non-Gaussian part of the fourth-order correlator of the  carrier density fluctuations reads
\begin{equation}\label{eq:n1_n4NG}
\begin{aligned}
&\braket{\Delta n(t_1)\Delta n(t_2)\Delta n(t_3)\Delta n(t_4)}_{\rm NG}=\frac{1}{L_x^4 L_y^4}\sum^{M_{\rm T}}_{i=1}\frac{z_i^4}{d^4}[f_{i}^2(1-f_{i})^2\\
&\times B(\gamma_i;t_1,t_2,t_3,t_4)+f_{i}(1-f_{i_1})(1-2f_{i_1})^2C(\gamma_i;t_1,t_2,t_3,t_4)]~,
\end{aligned}
\end{equation}
where
\begin{equation}\label{eq:Bdef}
\begin{aligned}
B(\gamma;t_1,t_2,t_3,t_4)&=
\sum_{\{a,b,c,d\}\in S_4} \Bigg\{
\Theta(t_a-t_b) e^{-\gamma (t_a-t_b)} \\
&\times \Big[\Theta(t_b-t_c)-\frac{1}{2}\Big] \Theta(t_c-t_d) e^{-\gamma (t_c-t_d)}\Bigg\}~,
\end{aligned}
\end{equation}

\begin{equation}\label{eq:Cdef}
\begin{aligned}
C(\gamma;t_1,t_2,t_3,t_4)&=\sum_{ \{a,b,c,d\} \in S_4} \Bigg\{ \Theta(t_a-t_b)\Theta(t_b-t_c) \\
&\times \Theta(t_c-t_d)e^{-\gamma (t_a-t_d)}\Bigg\}~,
\end{aligned}
\end{equation}
here, we sum $\{a,b,c,d\} $ over the $4!$ permutation of the array $\{1,2,3,4\}$, and $\Theta(x)$ denotes the Heaviside step function.
First of all, by making use of Eq.~\eqref{eq:n1_n4G}, we write the Gaussian contribution  $S_{n}^{(2){\rm G}}(\omega|\omega_0,\Delta \omega)$ as
\begin{equation}
\label{eq:S2_G}
\begin{aligned}
&S_{n}^{(2){\rm G}}(\omega|\omega_0,\Delta \omega)=  \lim_{T_t \to \infty}
\frac{8}{T_t} 
\int^{T_t/2}_{-T_t/2} d \tau_1 
\int^{T_t/2}_{-T_t/2} d \tau_2 e^{i \omega (\tau_1-\tau_2)}\\
&\times \frac{1}{T_s^4}   
\int^{\tau_1+T_s/2}_{\tau_1-T_s/2}
d t_1 
\int^{\tau_1+T_s/2}_{\tau_1-T_s/2} d t_2 
\int^{\tau_2+T_s/2}_{\tau_2-T_s/2}
d t_3 \int^{\tau_2+T_s/2}_{\tau_2-T_s/2} d t_4 \\
&\times  e^{i \omega_0  (t_1-t_2+t_3-t_4)} 
[\braket{\Delta n(t_1)\Delta n(t_3)}\braket{\Delta n(t_2)\Delta n(t_4)}\\
&+\braket{\Delta n(t_1)\Delta n(t_4)}\braket{\Delta n(t_2)\Delta n(t_3)}]~.
\end{aligned}
\end{equation}
After algebraic manipulations, the Gaussian contribution of the second spectrum reads
 \begin{equation}\label{eq:S2n_Gv2}
 \begin{aligned}
S_{n}^{(2){\rm G}}(\omega|\omega_0,\Delta \omega)&=4
\int^\infty_{-\infty} \frac{d \Omega }{2 \pi} {\cal F}(\Omega,\omega,\omega_0)\\
&\times S_{n}(\Omega+\omega/2)S_{n}(\Omega-\omega/2),
\end{aligned}
\end{equation}
in terms of the usual power spectrum of the carrier density fluctuations
\begin{equation}
 \begin{aligned}
S_n(\omega)&=\lim_{T_t \to \infty}\frac{1}{T_t} 
\int^{T_t/2}_{-T_t/2} d \tau e^{i \omega \tau} 
\int^{T_t/2}_{-T_t/2} d \tau' e^{-i \omega \tau'} \braket{\Delta n(\tau)\Delta n(\tau')}\\
&= \frac{1}{L_x^2 L_y^2} 
\sum^{M_{\rm T}}_{i=1} \frac{z_i^2}{d^2}f_{i}(1-f_{i})\frac{2 \gamma_i}{\omega^2+\gamma_i^2}
 \end{aligned} 
\end{equation}
and the function
\begin{equation}
 \begin{aligned}
 {\cal F}(\Omega,\omega,\omega_0)&=  [ g_s(\Omega+\omega/2-\omega_0)g_s(\Omega-\omega/2-\omega_0) \\
  &+g_s(\Omega+\omega/2+\omega_0) g_s(\Omega-\omega/2+\omega_0)]^2~,
 \end{aligned} 
\end{equation}
where $g_s(x)=\sin(\pi x/\Delta \omega)/(\pi x/\Delta \omega)$.

The non-Gaussian term of the second spectrum $S_{n}^{(2){\rm NG}}(\omega|\omega_0,\Delta \omega)=S_{n}^{(2)}(\omega|\omega_0,\Delta \omega)-S_{n}^{(2){\rm G}}(\omega|\omega_0,\Delta \omega)$, is obtained by using the fourth-order correlator of the carrier density fluctuations, Eq.~\eqref{eq:n1_n4NG},
\begin{equation}\label{eq:S2NG_v0}
\begin{aligned}
&S_{n}^{(2){\rm NG}}(\omega|\omega_0,\Delta \omega)=8 \int^{\infty}_{-\infty}\frac{d \Omega}{2\pi}\int^{\infty}_{-\infty}\frac{d \Omega'}{2\pi}g_s(\omega_0-\Omega)   \\  
&\times g_s(\omega_0+\omega-\Omega) g_s(\omega_0-\Omega')   g_s(\omega_0-\omega-\Omega') \\
&\times \frac{1}{L_x^4 L_y^4}\ \sum^{M_{\rm T}}_{i=1}  \frac{z_i^4}{d^4}[ f_i^2(1-f_i)^2 \tilde{B}(\gamma_i;\Omega,\omega-\Omega,\Omega',-\omega-\Omega')\\
&+ f_i(1-f_i)
(1-2f_i)^2 \tilde{C}(\gamma_i;\Omega,\omega-\Omega,\Omega',-\omega-\Omega')]~,
\end{aligned}
\end{equation}
where
\begin{equation}
\begin{aligned}
 \tilde{B}(\gamma_i;\Omega_1,\Omega_2,\Omega_3,\Omega_4)&=
\prod^4_{i=1} \int_{-\infty}^\infty d t_i e^{i \Omega_i t_i} B(\gamma_i;t_1,t_2,t_3,t_4)~,\\
 \tilde{C}(\gamma_i;\Omega_1,\Omega_2,\Omega_3,\Omega_4)&=
\prod^4_{i=1} \int_{-\infty}^\infty d t_i e^{i \Omega_i t_i} C(\gamma_i;t_1,t_2,t_3,t_4)~.
\end{aligned}
\end{equation}
Eq.~(\ref{eq:S2NG_v0}) is the first important result of this work. Together with Eq.~ (\ref{eq:S2n_Gv2}), it provides the exact expression of the second spectrum of charge carriers fluctuations in graphene due to trapping/detrapping processes valid for any distribution of tunneling rates and energies of the trap states,  which are dependent on the substrate status.  

\emph{Homogeneous traps distribution}. We introduce a set of reasonable and general assumptions. 
First, we assume that the stationary probability corresponds to the equilibrium occupation function such that $f_i=f_{\rm D}(\epsilon_i)$, where $f_{\rm D}(x)=1/[e^{(x-\mu_0)/(k_{\rm B}T)}+1]$, $\mu_0$ is the Fermi level, and $k_{\rm B}T$ is the thermal energy.
Second, we assume that traps are homogeneously distributed in the substrate so that the number of trap states per unit volume and energy $ {\cal D}(\bm{r},z,\epsilon)=\sum^{M_{\rm T}}_{i=1}\delta(\bm{r}-\bm{r}_i) \delta(z-z_i)  \delta(\varepsilon-\varepsilon_i) \to{\cal D}(\epsilon) $. 
Finally, we assume that the switching rates depend only on the trap distance from the graphene layer as $\gamma(z)=\gamma_M e^{-|z-d|/\ell_0}$, where the typical width of the substrate is $d\sim 100~{\rm nm}$, and usually the tunneling parameters are $\gamma_0 \sim 10^{10}~{\rm s}^{-1}$ and $\ell_0 \sim 1-20$~\AA, respectively~\cite{hsu_1970,balandin_natnano_2013}.
According to the monotonic $z$-dependence of the switching rates,  the maximum and minimum rates are respectively equal to  $\gamma_M$ and  $\gamma_m=\gamma_M e^{-d/\ell_0}$. 
We focus on the experimentally relevant regime $\omega \ll \Delta \omega \ll \omega_0$, and we will consider that both frequencies $\omega$ and $\omega_0$ are smaller (larger) than the maximum rate $\gamma_M$ (minimum rate $\gamma_m$). Moreover, we will call fast (slow) fluctuators the charge traps with tunneling rates $\gamma$ larger (smaller) than the bandwidth $\Delta \omega$, i.e. $\Delta \omega <\gamma<\gamma_M$ ($\gamma_m <\gamma<\Delta \omega$),  and we will neglect corrections of order $(\ell_0/d)\ln[\gamma_M/(\Delta \omega)]\ll 1$. 
Following the above assumptions, the Gaussian part of the second spectrum 
$S_{n}^{(2){\rm G}}(\omega|\omega_0,\Delta \omega)$ is expressed as
\begin{equation}\label{eq:S2n_G_final}
 S_{n}^{(2){\rm G}}(\omega|\omega_0,\Delta \omega)=   \frac{ F_0^2 \ell_0^2 }{L_x^2 L_y^2} \Big[ \frac{8\pi \Delta \omega}{3 \omega_0^2}+\frac{d}{\ell_0}
\frac{4(\Delta \omega)^4}{ \pi^3 \omega_0^4} \frac{1}{\omega}
 \Big]~,
\end{equation}
where
\begin{equation}\label{eq:F0}
F_0=\int^\Lambda_{-\Lambda} d \epsilon  {\cal D}(\epsilon) f_{\rm D}(\epsilon)[1-f_{\rm D}(\epsilon)]~.  
\end{equation}
Here, the white term (i.e. $\omega$-independent) is proportional to the frequency bandwidth $\Delta \omega$ and it originates exclusively from the action of the fast fluctuators, while the $1/\omega$ dependence is due entirely to slow fluctuators.
The independence of $\omega$ of the Gaussian part of the second spectrum is a general result~\cite{weissman_rmp_1988}, as long as the power spectrum has no structure on the scale of the measurement bandwidth~$\Delta \omega$, as it is the case for the contribution of $S_n(\omega)$ due to the fast fluctuators.
On the other hand, the contribution of $S_n(\omega)$ related to the slow fluctuators appears as the envelope of contributions of single traps which are very sharply peaked. So, it has a nontrivial structure on the scale of the measurement bandwidth $\Delta \omega$, and it leads to an $\omega$-dependent second spectrum.
By comparing these two contributions of the Gaussian part of the second spectrum, we obtain the ratio between the $1/f$ and the white contributions as $\sim [3/(2\pi^4)](d/\ell_0) (\Delta \omega/\omega_0)^3 \omega_0/\omega$, 
and it is clear that the determination of which term prevails over the other depends on the specifics of the experiment. 
For instance, with a substrate of width $d=100$~nm, a tunneling parameter $\ell_0=1$~\AA, and a frequency bandwidth $\Delta \omega=\omega_0/10$, 
the $1/\omega$ term is dominant for $\omega/\omega_0 \ll 3/(2\pi^4) \approx 10^{-2}$.

Moreover, by enforcing in Eq.~\eqref{eq:S2NG_v0} the assumptions of homogeneous traps distribution, the non-Gaussian part of the second spectrum reads
\begin{equation}\label{eq:S2n_NG_final}
S_{n}^{(2){\rm NG}}(\omega|\omega_0,\Delta \omega)\approx \frac{\ell_0}{L_x^3 L_y^3} \bigg[ 
\frac{4 (\Delta \omega)^2 F_2}{\pi \omega_0^3} 
+\frac{3 (\Delta \omega)^3}{2 \pi^3 \omega_0^3} \frac{3 F_2-F_0}{\omega} 
\bigg]~,
\end{equation}
where
\begin{equation}\label{eq:F2}
F_2=\int^\Lambda_{-\Lambda} d \epsilon  {\cal D}(\epsilon) f_{\rm D}(\epsilon)[1-f_{\rm D}(\epsilon)][1-2 f_{\rm D}(\epsilon)]^2~.  
\end{equation}
We observe that the non-Gaussian second spectrum does not depend on the width of the substrate $d$, contrary to the Gaussian contribution in Eq.~\eqref{eq:S2n_G_final}.
Moreover, for a given frequency bandwidth $\Delta \omega=\omega_0/10$, the $1/\omega$  term is the leading contribution of $S_{n}^{(2){\rm NG}}(\omega|\omega_0,\Delta \omega)$ provided that $\omega/\omega_0 \ll 10^{-2} (3 F_2-F_0)  /F_2$.
All details for the derivations of Eq.~\eqref{eq:S2n_G_final}  and Eq.~\eqref{eq:S2n_NG_final} are included in the supplemental material.

\begin{figure}[t]
\centering
\includegraphics[width=\columnwidth]{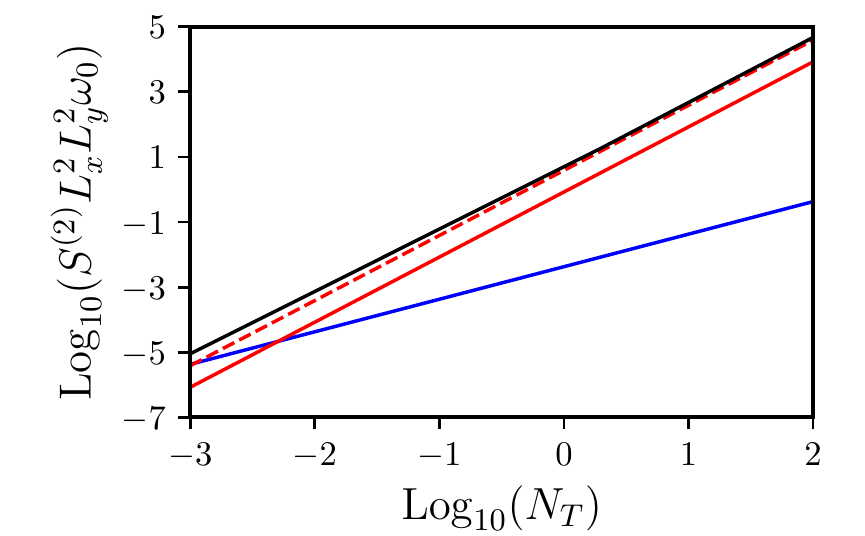}
\caption{
Second spectrum $S^{(2)}(\omega|\omega_0,\Delta \omega)$, in units of $1/(L_x^2L_y^2 \omega_0)$,  as a function of $N_T={\cal D}(\mu_0) k_{\rm B} T \ell_0 L_x L_y$, in a Log-Log scale. 
Here, the red lines represent respectively the white spectrum term (solid line) and the $1/f$ spectrum term (dashed line) of $S^{2{\rm G}}(\omega|\omega_0,\Delta \omega)$, defined in Eq.~\eqref{eq:S2n_G_thermal}. The blue line corresponds to the white spectrum term of $S^{(2){\rm NG}}(\omega|\omega_0,\Delta \omega)$, defined in Eq.~\eqref{eq:S2n_NG_thermal}. The black line denotes the sum of all contributions. We have set $\Delta \omega/\omega_0=10^{-1}$, $\omega/\omega_0=10^{-2}$, and $d/\ell_0=3 \times 10^3$.
\label{fig:thermal}}
\end{figure}

\emph{Low temperature dependencies}. The temperature dependence of the second spectrum of carrier density fluctuations enters via the functions $F_0$ and $F_2$, defined in Eqs~\eqref{eq:F0} and \eqref{eq:F2}, which also depend on the energy distribution of trap states.
By assuming that the density of trap states ${\cal D}(\epsilon)$ is a smooth function in the thermal energy window around the Fermi level $\mu_0$, we can write respectively $F_0 \approx {\cal D}(\mu_0)k_{\rm B} T+(\pi^2/12) \partial^2_\epsilon{\cal D}(\epsilon)|_{\epsilon=\mu_0} (k_{\rm B} T)^3$ and $F_2 \approx F_0/3 + (2/3) \partial^2_\epsilon{\cal D}(\epsilon)|_{\epsilon=\mu_0} (k_{\rm B} T)^3$.
Thus in the low-temperature regime Eqs.~\eqref{eq:S2n_G_final} and \eqref{eq:S2n_NG_final} take the following form
\begin{equation}\label{eq:S2n_G_thermal}
 S_{n}^{(2){\rm G}}(\omega|\omega_0,\Delta \omega)=   \frac{  {\cal D}^2(\mu_0) (k_{\rm B} T)^2 \ell_0^2 }{L_x^2 L_y^2} \Big[ \frac{8\pi \Delta \omega}{3 \omega_0^2}+\frac{d}{\ell_0}
\frac{4(\Delta \omega)^4}{ \pi^3 \omega_0^4} \frac{1}{\omega}
 \Big]~,
\end{equation}
\begin{equation}\label{eq:S2n_NG_thermal}
\begin{aligned}
S_{n}^{(2){\rm NG}}(\omega|\omega_0,\Delta \omega)&\approx \frac{\ell_0}{L_x^3 L_y^3} \bigg[ 
\frac{4 (\Delta \omega)^2 {\cal D}(\mu_0)k_{\rm B} T}{3 \pi \omega_0^3} 
\\
&+\frac{3 (\Delta \omega)^4}{ \pi^3 \omega_0^4} \frac{\partial^2_\epsilon{\cal D}(\epsilon)|_{\epsilon=\mu_0} (k_{\rm B} T)^3}{\omega} 
\bigg]  ~.  
\end{aligned}
\end{equation}
The Gaussian contribution exhibits a quadratic temperature dependence. On the other hand, in the non-Gaussian part, the white term displays a linear temperature dependence whereas the $1/\omega$ term scales as $T^3$. 
It is comparable to the white term only for very small frequencies $\omega$ such that $\omega/\omega_0  \lesssim 10^{-2}   (k_{\rm B} T)^2 \partial^2_\epsilon{\cal D}(\epsilon)|_{\epsilon=\mu_0}/{\cal D}(\mu_0)$.
Therefore, in the low-temperature regime, the second spectrum of charge carrier fluctuations may display a $1/f$ behavior essentially due to the Gaussian contribution,  the non-Gaussian $1/f$ term being strongly suppressed.
Indeed, the ratio between the  $1/f$ terms of the non-Gaussian and the Gaussian parts goes as $\sim \partial^2_\epsilon{\cal D}(\epsilon)|_{\epsilon=\mu_0} k^2_{\rm B} T^2/[{\cal D}(\mu_0) d L_x L_y] $ whereas the the ratio between the $\omega$-independent terms goes as $\sim (\omega_0/\Delta \omega)/[{\cal D}(\mu_0) k_{\rm B} T \ell_0 L_x L_y]$. Therefore,  thanks to its linear temperature dependence, the white contribution of the non-Gaussian part of the second noise spectrum can be comparable to the usually dominant Gaussian part. 
This is illustrated in Fig.~\ref{fig:thermal}, where  $\omega$-independent term (red solid line) and the $1/\omega$ term (red dashed line) of  $S^{(2){\rm G}}(\omega|\omega_0,\Delta \omega)$ are compared with the white contribution of $S^{(2){\rm NG}}(\omega|\omega_0,\Delta \omega)$ (blue line), as a function of $N_T={\cal D}(\mu_0) k_{\rm B} T \ell_0 L_x L_y$, the number of thermally  activated charge trap states in an extremely thin slab of width $\ell_0$ (the total number of thermally activated trap states in the entire substrate is $d N_T/\ell_0$). The black solid line is the overall second spectrum.
Note that the $1/f$ contribution to the non-Gaussian part of the second spectrum is negligible for two reasons: we have assumed a smooth density of trap states ${\cal D}(\epsilon)$ and we are studying the low-temperature regime.
In Fig.~\ref{fig:thermal} we fixed $\Delta \omega/\omega_0=10^{-1}$ and  $\omega/\omega_0=10^{-2}$. In a substrate of width $d=300~{\rm nm}$ and a tunneling length $\ell_0=1$~\AA, the $1/f$ contribution of the Gaussian part is the leading term of $S^{(2)}(\omega|\omega_0,\Delta \omega)$. Furthermore, at extremely low temperatures, such that the total number of thermally activated trap states $d N_T/\ell_0$ is of the order of tens,  the white non-Gaussian part of the second spectrum becomes  larger than the white Gaussian part. Indeed, at extremely low temperatures, the complete second spectrum deviates from the quadratic temperature dependence of the Gaussian part. 

%
In summary, we have investigated the second spectrum of charge density fluctuations of a 2D electron gas by using a minimal approach based on the McWhorter model~\cite{mcwhorter_1957,pellegrino_epjst_2021}. In particular, the reference physical system is   a graphene monolayer in the large doping regime ($n\approx 10^{12}~{\rm cm}^{-12}$). 
Assuming that  charge traps are uniformly distributed in the substrate and in thermal equilibrium, we found in analytic form both the Gaussian and non-Gaussian parts of the second spectrum of carrier density fluctuations. The contribution due to  Gaussian fluctuations consists of a white and a $1/f$ term. The former is due to charge traps characterized by  switching rates $\gamma$ larger than the measurement bandwidth~$\Delta \omega$ (fast fluctuators), while the latter originates from charge traps with switching rates $\gamma$ smaller than ~$\Delta \omega$  (slow fluctuators). Slow fluctuators are responsible for a carrier density power spectrum with a structure on the  scale of the measurement bandwidth~$\Delta \omega$.
As a consequence, an additional $1/f$ contribution sums up to the expected frequency-independent term in the Gaussian part of the second spectrum~\cite{weissman_rmp_1988}.
In the case of a thick substrate, the number of slow fluctuators is much larger than fast fluctuators, 
thus there is a range of small $\omega/\omega_0$ where detection of the leading, $1/f$ part of the Gaussian second spectrum might be feasible. 
Also, the second spectrum of the non-Gaussian carrier density fluctuations consists of a $\omega$-independent term due to the action of the slow fluctuators and a $1/f$ term related to the fast fluctuators.
In general, the non-Gaussian contribution is negligible with respect to the Gaussian part. However, we demonstrated that in the low-temperature limit, the white contribution is the leading term of the non-Gaussian part of the second spectrum and it has a linear temperature dependence, contrary to the Gaussian part of the second spectrum which appears as a quadratic temperature function. Therefore, for extremely small $\omega/\omega_0$, the non-Gaussian part can become relevant.

\vspace{1em}

See the supplementary material for details of the derivations of the multi-time correlation functions and of the second spectrum of carrier density fluctuations under the assumption of a homogeneous traps distribution.

\vspace{1em}
The authors thank G. G. N. Angilella,  R. Fazio, and P. Hakonen for illuminating discussions and fruitful comments on various stages of this work. This research was supported by PNRR MUR project PE0000023-NQSTI,  COST Action CA21144 superqumap, and the Universit\`a degli Studi di Catania, Piano di Incentivi per la Ricerca di Ateneo 2020/2022 (progetto QUAPHENE and progetto Q-ICT).

\nocite{*}
\bibliography{02_literature}

\appendix
\clearpage 
\setcounter{section}{0}
\setcounter{equation}{0}%
\setcounter{figure}{0}%
\setcounter{table}{0}%

\setcounter{page}{1}

\renewcommand{\thetable}{S\arabic{table}}
\renewcommand{\theequation}{S\arabic{equation}}
\renewcommand{\thefigure}{S\arabic{figure}}
\renewcommand{\bibnumfmt}[1]{[S#1]}

\onecolumngrid

\begin{center}
 \textbf{\Large Supplemental Material for ``Second spectrum of charge carrier density fluctuations in graphene due to trapping/detrapping processes''}
\end{center}

\begin{center}
Francesco M.D. Pellegrino,$^{1\,,2\,,3}$
Giuseppe Falci,$^{1\,,2\,,3}$ and
Elisabetta Paladino,$^{1\,,2\,,3}$
\end{center}

\begin{center}
{\small
$^1$\!{\it Dipartimento di Fisica e Astronomia ``Ettore Majorana'', Universit\`a di Catania, Via S. Sofia 64, I-95123 Catania,~Italy.}

$^2$\!{\it INFN, Sez.~Catania, I-95123 Catania,~Italy.}

$^3$\!{\it CNR-IMM, Via S. Sofia 64, I-95123 Catania,~Italy.}

}
\end{center}

\twocolumngrid


\section*{Multi-time correlation functions}

Within the McWhorther model~\cite{kogan_book,pellegrino_commphys_2020,pellegrino_jstat_2019},
the probability of the trap $i$ having an occupation number $X(i, t)$ at time $t$, given that the trap $j$ had a specific occupation number $X(j, t_0)$ at time $t_0$ is expressed as
\begin{equation}\label{eq:conditional}
\begin{aligned}
P[X(i,t)|X(j,t_0)]&= \delta_{i, j} p_i[X(i, t)|X(i, t_0)](t-t_0)\\
&+ w_i (X(i, t))  ~,
\end{aligned}
 \end{equation}
where $w_i(X(i, t))$ is the stationary probability of trap $i$, it depends on $X(i, t)$ as  $w_i(1)=f_{i}$ and $w_i(0)=1-f_{i}$,  and different traps are uncorrelated.
The matrix $p_i$ is expressed as
\begin{equation}
\begin{bmatrix}
 p_i[0|0](t) &  p_i[0|1](t)  \\
 p_i[1|0](t) &  p_i[1|1](t)  \\
\end{bmatrix}=
\begin{bmatrix}
f_i & -(1-f_i) \\
- f_i   & 1-f_i  
\end{bmatrix} e^{-\gamma_i t}~,
\end{equation}
where $t\ge0$, and  $\gamma_i$ is the switching rate between the two states of the stochastic process.
Thus, we evaluate explicitly the second-order correlator as
\begin{equation}\label{eq:x1_x2}
\begin{aligned}
&\braket{x(i_1,t_1)x(i_2,t_2)}\\
&=\delta_{i_1,i_2}[(1-f_{i_1})^2 (p_{i_1}[1,1](t_1-t_2)+ w_{i_1}(1) )w_{i_1}(1) \\
&-f_{i_1} (1-f_{i_1})(p_{i_1}[0,1](t_1-t_2)+ w_{i_1}(0) )w_{i_1}(1) \\
&- (1-f_{i_1})f_{i_1} (p_{i_1}[1,0](t_1-t_2)+ w_{i_1}(1) )w_{i_1}(0)  \\
&+f^2_{i_1} (p_{i_1}[0,0](t_1-t_2)+ w_{i_1}(0) )w_{i_1}(0)  
]\\
&=\delta_{i_1,i_2}f_{i_1}(1-f_{i_1})e^{-\gamma_{i_1}|t_1-t_2|}~,
\end{aligned}
\end{equation}
where $x(i,t)=X(i,t)-f_{i}$.
Exploiting Markovianity~\cite{kogan_book} 
\begin{equation}\label{eq:condchain}
\begin{aligned}
& P[X(i_N, t_N)|X(i_{N-1}, t_{N-1});\ldots ;X(i_0, t_0)] \\      &=P[X(i_N, t_N)|X(i_{N-1}, t_{N-1})]~,    
\end{aligned}
\end{equation}
where the times are chronologically ordered, i.e. $t_0 \le t_1 \le \ldots \le t_N$,
we can calculate any multi-time correlator in terms of the two-time conditional probability, Eq.\eqref{eq:conditional}.  
In particular, 
the fourth-order multi-time correlator 
reads
\begin{equation}\label{eq:x1_x4}
\begin{aligned}
&\braket{x(i_1,t_1)x(i_2,t_2)x(i_3,t_3)x(i_4,t_4)}=\\
&=\braket{x(i_1,t_1)x(i_2,t_2)x(i_3,t_3)x(i_4,t_4)}_{\rm G}\\
&+\braket{x(i_1,t_1)x(i_2,t_2)x(i_3,t_3)x(i_4,t_4)}_{\rm NG},
\end{aligned}
\end{equation}
where the first contribution is the Gaussian term
\begin{equation}\label{eq:x1_x4G}
\begin{aligned}
&\braket{x(i_1,t_1)x(i_2,t_2)x(i_3,t_3)x(i_4,t_4)}_{\rm G}=\\
&=\delta_{i_1,i_2}\delta_{i_3,i_4}\braket{x(i_1,t_1)x(i_1,t_2)}\braket{x(i_3,t_3)x(i_3,t_4)}\\
&+\delta_{i_1,i_3}\delta_{i_2,i_4}\braket{x(i_1,t_1)x(i_1,t_3)}\braket{x(i_2,t_2)x(i_2,t_4)}\\
&+\delta_{i_1,i_4}\delta_{i_2,i_3}\braket{x(i_1,t_1)x(i_1,t_4)}\braket{x(i_2,t_2)x(i_2,t_3)}~,
\end{aligned}
\end{equation}
%
and it is expressed in terms of the second-order correlators,  while  the remaining contribution denotes the non-Gaussian term 
\begin{equation}\label{eq:x1_x4NG}
\begin{aligned}
&\braket{x(i_1,t_1)x(i_2,t_2)x(i_3,t_3)x(i_4,t_4)}_{\rm NG}=\\
&=\delta_{i_1,i_2}\delta_{i_2,i_3}\delta_{i_3,i_4}[f_{i_1}^2(1-f_{i_1})^2B(\gamma_{i_1};t_1,t_2,t_3,t_4)\\
&+f_{i_1}(1-f_{i_1})(1-2f_{i_1})^2C(\gamma_{i_1};t_1,t_2,t_3,t_4)],
\end{aligned}
\end{equation}
where $B(\gamma;t_1,t_2,t_3,t_4)$ and $C(\gamma;t_1,t_2,t_3,t_4)$ are defined in the main text.

\section*{Gaussian part of the second spectrum of charge carrier density fluctuations}

In this Section, we derive the Gaussian part of the second spectrum of carrier density fluctuations under the assumption of a homogeneous traps distribution.
It is convenient to split 
$S_{n}^{(2){\rm G}}(\omega|\omega_0,\Delta \omega)$ into four parts, as follows 
\begin{equation}
\begin{aligned}
S_{n}^{(2){\rm G}}(\omega|\omega_0,\Delta \omega)&=
\sum_{\ell=\{{\rm f,s}\}} \sum_{\ell'=\{{\rm f,s}\}} S_{n}^{(2){\rm G}\ell\ell'}(\omega|\omega_0,\Delta \omega)~,\\
S_{n}^{(2){\rm G}\ell\ell'}(\omega|\omega_0,\Delta \omega)&=4
\int^\infty_{-\infty} \frac{d \Omega }{2 \pi} {\cal F}(\Omega,\omega,\omega_0)\\
&\times S^\ell_{n}(\Omega+\omega/2)S^{\ell'}_{n}(\Omega-\omega/2)~,
\end{aligned}
\end{equation}
which is obtained by decomposing the power spectrum of the carrier density fluctuations as $S_n(\omega)=S^{\rm s}_n(\omega)+S^{\rm f}_n(\omega)$, where the first term is related to the slow fluctuators as 
\begin{equation}
 \begin{aligned}
S^{\rm s}_n(\omega)&= \frac{F_0 \ell_0 }{L_x L_y} \int_{\gamma_m}^{\Delta \omega} d \gamma
\frac{[d+\ell_0 \ln(\gamma/\gamma_M)]^2}{d^2}\frac{2}{\omega^2+\gamma^2}~,
 \end{aligned} 
\end{equation}
 $F_0$ is defined in the main text,
%
%
and the second contribution is related to the fast fluctuators as
\begin{equation}
 \begin{aligned}
S^{\rm f}_n(\omega)&= \frac{F_0 \ell_0 }{L_x L_y} \int^{\gamma_M}_{\Delta \omega} d \gamma
\frac{[d+\ell_0 \ln(\gamma/\gamma_M)]^2}{d^2}\frac{2}{\omega^2+\gamma^2}\\
&\approx \frac{F_0 \ell_0 }{L_x L_y} \int^{\gamma_M}_{\Delta \omega} d \gamma \frac{2}{\omega^2+\gamma^2}\\
&= \frac{2 F_0 \ell_0 }{L_x L_y \omega}[\arctan(\gamma_M/\omega)-\arctan(\Delta \omega/\omega)]~,
 \end{aligned} 
\end{equation}
where we have neglected 
terms of order $(\ell_0/d)\ln[\gamma_M/(\Delta \omega)]\ll 1$.
The contribution which involves only  fast fluctuators reads
\begin{equation}
\begin{aligned}
& S_{n}^{(2){\rm G ff}}(\omega|\omega_0,\Delta \omega)=\frac{4 F_0^2 \ell_0 ^2}{L_x^2 L_y^2}
\int^\infty_{-\infty} \frac{d \Omega }{2 \pi}  \int^{\gamma_M}_{\Delta \omega} d \gamma \int^{\gamma_M}_{\Delta \omega} d \gamma'
\\
&\times   \frac{[d+\ell_0 \ln(\gamma/\gamma_M)]^2}{d^2} \frac{[d+\ell_0 \ln(\gamma'/\gamma_M)]^2}{d^2} 
{\cal F}(\Omega,\omega,\omega_0)
\\
&\times \frac{2}{(\Omega+\omega/2)^2+\gamma^2}
 \frac{2}{(\Omega-\omega/2)^2+{\gamma'}^2}~\\
&\approx \frac{4 F_0^2 \ell_0^2 }{L_x^2 L_y^2}
\int^\infty_{-\infty} \frac{d \Omega }{2 \pi}  \int^{\gamma_M}_{\Delta \omega} d \gamma \int^{\gamma_M}_{\Delta \omega} d \gamma' {\cal F}(\Omega,\omega,\omega_0)
\\
&\times  \frac{2}{(\Omega+\omega/2)^2+\gamma^2}
 \frac{2}{(\Omega-\omega/2)^2+{\gamma'}^2}~.
\end{aligned}
\end{equation}
Here, since we are interested in the regime $\omega\ll \Delta \omega \ll \omega_0$, we 
approximate ${\cal F}(\Omega,\omega,\omega_0)\approx {\cal F}(\Omega,0,\omega_0)$. Moreover, since fast fluctuators have tunneling rates such that $\gamma,\gamma'\gg \Delta \omega$, we can write
\begin{equation}\label{eq:S2gff}
\begin{aligned}
& S_{n}^{(2){\rm G ff}}(\omega|\omega_0,\Delta \omega)\approx \frac{4 F_0^2 \ell_0^2 }{L_x^2 L_y^2}
\int^\infty_{-\infty} \frac{d \Omega }{2 \pi}  {\cal F}(\Omega,0,\omega_0) \int^{\gamma_M}_{\Delta \omega} d \gamma
\\
&\times    \frac{2}{\omega_0^2+\gamma^2} \int^{\gamma_M}_{\Delta \omega} d \gamma' 
 \frac{2}{\omega_0^2+{\gamma'}^2}=
 \frac{8 \Delta \omega}{3\pi}  \left[S^{\rm f}_n(\omega_0)\right]^2\\
 & \approx \frac{8 F_0^2 \ell_0^2 }{L_x^2 L_y^2} \frac{\pi \Delta \omega}{3 \omega_0^2}~,
\end{aligned}
\end{equation}
in the last line we used that $S^{\rm f}_n(\omega_0)\approx \pi F_0 \ell_0/(L_x L_y \omega_0)$, since $ \Delta \omega \ll  \omega_0 \ll\gamma_M$.
%
We now face the evaluation of the Gaussian terms of the second spectrum which involves both slow and fast fluctuators, i.e. $S_{n}^{(2){\rm G fs}}(\omega|\omega_0,\Delta \omega)$ and $S_{n}^{(2){\rm G sf}}(\omega|\omega_0,\Delta \omega)$.
Starting from their definitions, one has $S_{n}^{(2){\rm G sf}}(\omega|\omega_0,\Delta \omega)=S_{n}^{(2){\rm G fs}}(-\omega|\omega_0,\Delta \omega)$. By using that ${\cal F}(\Omega,\omega,\omega_0)={\cal F}(\Omega,-\omega,\omega_0)$, we straightforwardly find $S_{n}^{(2){\rm G fs}}(\omega|\omega_0,\Delta \omega)=S_{n}^{(2){\rm G fs}}(-\omega|\omega_0,\Delta \omega)$, which leads to  $S_{n}^{(2){\rm G sf}}(\omega|\omega_0,\Delta \omega)=S_{n}^{(2){\rm G fs}}(\omega|\omega_0,\Delta \omega)$.
Therefore, it is enough to calculate explicitly only $S_{n}^{(2){\rm G fs}}(\omega|\omega_0,\Delta \omega)$ as
\begin{equation}
\begin{aligned}
& S_{n}^{(2){\rm G fs}}(\omega|\omega_0,\Delta \omega)=\frac{4 F_0^2 \ell_0 ^2}{L_x^2 L_y^2}
\int^\infty_{-\infty} \frac{d \Omega }{2 \pi}  \int^{\gamma_M}_{\Delta \omega} d \gamma \int_{\gamma_m}^{\Delta \omega} d \gamma'
\\
&\times   \frac{[d+\ell_0 \ln(\gamma/\gamma_M)]^2}{d^2} \frac{[d+\ell_0 \ln(\gamma'/\gamma_M)]^2}{d^2} 
{\cal F}(\Omega,\omega,\omega_0)
\\
&\times \frac{2}{(\Omega+\omega/2)^2+\gamma^2}
 \frac{2}{(\Omega-\omega/2)^2+{\gamma'}^2}~\\
&\approx \frac{4 F_0^2 \ell_0 ^2}{L_x^2 L_y^2}
\int^\infty_{-\infty} \frac{d \Omega }{2 \pi}  \int^{\gamma_M}_{\Delta \omega} d \gamma \int_{\gamma_m}^{\Delta \omega} d \gamma'  \frac{[d+\ell_0 \ln(\gamma'/\gamma_M)]^2}{d^2} 
\\
&\times  
{\cal F}(\Omega,\omega,\omega_0) \frac{2}{(\Omega+\omega/2)^2+\gamma^2}
 \frac{2}{(\Omega-\omega/2)^2+{\gamma'}^2},
\end{aligned}
\end{equation}
here, since $\gamma'\ll \Delta \omega$, we  approximate $\gamma'/[(\Omega-\omega/2)^2+\gamma'^2]\approx \pi \delta(\Omega-\omega/2)$, which is a Dirac delta function, such that
\begin{equation}\label{eq:S2Gfs_v0}
\begin{aligned}
 S_{n}^{(2){\rm G fs}}(\omega|\omega_0,\Delta \omega)
&\approx \frac{4 F_0^2 \ell_0 d}{L_x^2 L_y^2} {\cal F}(\omega/2,\omega,\omega_0) 
\frac{2}{3 \omega}
\\
&\times  
[\arctan(\gamma_M/\omega)-\arctan(\Delta \omega/\omega)]\\
&\times [1+(\ell_0/d)\ln(\Delta \omega/\gamma_M)]^3.
\end{aligned}
\end{equation}
%
%
In the regime  $\omega\ll \Delta \omega \ll \omega_0\ll \gamma_M$ a series of approximations can be applied.
Firstly, we use ${\cal F}(\omega/2,\omega,\omega_0)\approx {\cal F}(0,0,\omega_0)=4 [\sin(\pi \omega_0/\Delta \omega )/(\pi \omega_0/ \Delta \omega )]^4$. Here, the rapidly oscillating function $\sin^4(\pi \omega_0/\Delta \omega )$ appears, and we replace it with its average value $3/8$, such that ${\cal F}(\omega/2,\omega,\omega_0)\to 3[(\Delta \omega/(\pi \omega_0)]^4/2$.
Moreover we neglect the correction $(\ell_0/d)\ln[\gamma_M/(\Delta \omega)]\ll 1$.
Finally,  we take the leading term of the  expansion $[\arctan(\gamma_M/\omega)-\arctan(\Delta \omega/\omega)]\approx \omega/(\Delta \omega)$, which is valid for small $\omega$ with respect $\Delta \omega$ and $\gamma_M$.
Under these  approximations, the Gaussian part of the second spectrum which involves slow and fast fluctuators reads
\begin{equation}\label{eq:S2Gfs}
\begin{aligned}
& S_{n}^{(2){\rm G fs}}(\omega|\omega_0,\Delta \omega)
\approx \frac{4 F_0^2 \ell_0 d}{L_x^2 L_y^2} \frac{(\Delta \omega)^3}{\pi^4 \omega_0^4}~,
\end{aligned}
\end{equation}
which appears as a white spectrum with respect to  $\omega$, analogously to the term due only to fast fluctuators. 
We conclude the calculation of the Gaussian part of the second
spectrum with the term involving only slow fluctuators 
\begin{equation}
\begin{aligned}
& S_{n}^{(2){\rm G ss}}(\omega|\omega_0,\Delta \omega)=\frac{4 F_0^2 \ell_0 ^2}{L_x^2 L_y^2}
\int^\infty_{-\infty} \frac{d \Omega }{2 \pi}  \int_{\gamma_m}^{\Delta \omega} d \gamma \int_{\gamma_m}^{\Delta \omega} d \gamma'
\\
&\times   \frac{[d+\ell_0 \ln(\gamma/\gamma_M)]^2}{d^2} \frac{[d+\ell_0 \ln(\gamma'/\gamma_M)]^2}{d^2} 
{\cal F}(\Omega,\omega,\omega_0)\\
&\times \frac{2}{(\Omega+\omega/2)^2+\gamma^2}
 \frac{2}{(\Omega-\omega/2)^2+{\gamma'}^2}~.    
\end{aligned}
\end{equation}
By performing the integral over $\Omega$,  since $\gamma,\gamma'\ll \Delta \omega$, we approximate $\gamma/[(\Omega+\omega/2)^2+\gamma^2]\approx \pi \delta(\Omega+\omega/2)$ and $\gamma'/[(\Omega-\omega/2)^2+\gamma'^2]\approx \pi \delta(\Omega-\omega/2)$,   close to $\Omega=-\omega/2$ and $\Omega=\omega/2$ respectively.
Moreover, with the approximation indicated above Eq.~\eqref{eq:S2Gfs}, we obtain
\begin{equation}\label{eq:S2Gss}
\begin{aligned}
 S_{n}^{(2){\rm G ss}}(\omega|\omega_0,\Delta \omega)
 &\approx \frac{4 F_0^2 \ell_0 d}{L_x^2 L_y^2}
\frac{(\Delta \omega)^4}{\pi^4 \omega_0^4} \frac{2}{\omega} \arctan(\Delta \omega/\omega)\\
& \approx \frac{4 F_0^2 \ell_0 d}{L_x^2 L_y^2}
\frac{(\Delta \omega)^4}{\pi^4 \omega_0^4} \frac{2}{\omega} \Big[\frac{\pi}{2 }-\frac{\omega}{\Delta \omega}\Big]~. 
\end{aligned}
\end{equation}
The leading term of the Gaussian part of the second spectrum which involves only the slow fluctuators depends on $\omega$ and is a $1/f$ contribution.
The subdominant term in Eq.~\eqref{eq:S2Gss} is independent of $\omega$, and it cancels out the contribution $S_{n}^{(2){\rm G fs}}(\omega|\omega_0,\Delta  \omega)+S_{n}^{(2){\rm G sf}}(\omega|\omega_0,\Delta \omega)$, see Eq.~\eqref{eq:S2Gfs}.
Therefore, by summing up all the terms of $S_{n}^{(2){\rm G}}(\omega|\omega_0,\Delta \omega)$, we have
\begin{equation*}\label{eq:S2n_G_final_SM}
 S_{n}^{(2){\rm G}}(\omega|\omega_0,\Delta \omega)=   \frac{ F_0^2 \ell_0^2 }{L_x^2 L_y^2} \Big[ \frac{8\pi \Delta \omega}{3 \omega_0^2}+\frac{d}{\ell_0}
\frac{4(\Delta \omega)^4}{ \pi^3 \omega_0^4} \frac{1}{\omega}
 \Big]~.
\end{equation*}
%
%

\section*{Non-Gaussian part of the second spectrum of charge carrier density fluctuations}

In this Section, we derive the non-Gaussian part of the second spectrum of the carrier density fluctuations under the assumption of homogeneous traps distribution.
Analogously to the Gaussian part, here we decompose
the non-Gaussian contribution of the second spectrum as $S^{\rm NG}_n(\omega)=S^{\rm NGs}_n(\omega)+S^{\rm NGf}_n(\omega)$, where the latter term is related to fast fluctuators, while the former one originates from  slow fluctuators.
By enforcing the assumptions of homogeneous traps distribution,
we obtain
\begin{equation}\label{eq:S2NG_f}
\begin{aligned}
&S_{n}^{(2){\rm NG}\ell}(\omega|\omega_0,\Delta \omega)=8 \int^{\infty}_{-\infty}\frac{d \Omega}{2\pi}\int^{\infty}_{-\infty}\frac{d \Omega'}{2\pi}g_s(\omega_0-\Omega)   \\  
&\times g_s(\omega_0+\omega-\Omega) g_s(\omega_0-\Omega')   g_s(\omega_0-\omega-\Omega') \\
&\times\Big[\frac{F_0-F_2}{4}\tilde{G}^{\ell}_B(\Omega,\omega-\Omega,\Omega',-\omega-\Omega') \\
&+F_2 \tilde{G}^{ \ell}_C(\Omega,\omega-\Omega,\Omega',-\omega-\Omega')  \Big]~,
\end{aligned}
\end{equation}
where $\ell={\rm f,s}$, and $F_2$ is defined in the main-text.
%
%
Firstly, we explicitly calculate the contribution $S_{n}^{(2){\rm NGf}}(\omega|\omega_0,\Delta \omega)$ which involves the fast fluctuators, where we have
\begin{equation}
\begin{aligned}
\tilde{G}^{\rm f}_B (\Omega_1,\Omega_2,\Omega_3,\Omega_4)&=\frac{\ell_0}{L_x^3 L_y^3} \int^{\gamma_M}_{\Delta \omega } \frac{d \gamma}{\gamma } 
\frac{[d+\ell_0 \ln(\gamma'/\gamma_M)]^4}{d^4} \\
&\times\tilde{B}(\gamma; \Omega_1,\Omega_2,\Omega_3,\Omega_4)\\
&\approx \frac{\ell_0}{L_x^3 L_y^3} \int^{\gamma_M}_{\Delta \omega } \frac{d \gamma}{\gamma } \tilde{B}(\gamma; \Omega_1,\Omega_2,\Omega_3,\Omega_4)~,
\end{aligned}
\end{equation}
and
\begin{equation}
\begin{aligned}
\tilde{G}^{\rm f}_C (\Omega_1,\Omega_2,\Omega_3,\Omega_4)&=\frac{\ell_0}{L_x^3 L_y^3} \int^{\gamma_M}_{\Delta \omega } \frac{d \gamma}{\gamma } 
\frac{[d+\ell_0 \ln(\gamma'/\gamma_M)]^4}{d^4} \\
&\times\tilde{C}(\gamma; \Omega_1,\Omega_2,\Omega_3,\Omega_4)\\
&\approx \frac{\ell_0}{L_x^3 L_y^3} \int^{\gamma_M}_{\Delta \omega } \frac{d \gamma}{\gamma } \tilde{C}(\gamma; \Omega_1,\Omega_2,\Omega_3,\Omega_4)~,
\end{aligned}
\end{equation}
%
here we have neglected terms of order $(\ell_0/d)\ln[\gamma_M/(\Delta \omega)]\ll 1$.
Since we are interested in the regime $\omega\ll \Delta \omega \ll \omega_0$ and fast fluctuators have tunneling rates such that $\gamma\gg \Delta \omega$,  it is possible to approximate $S_{n}^{(2){\rm NGf}}(\omega|\omega_0,\Delta \omega)\approx S_{n}^{(2){\rm NGf}}(0|\omega_0,\Delta \omega)$. Moreover, since $\tilde{G}^{\rm f}_B(\Omega,-\Omega,\Omega',-\Omega') $ and $\tilde{G}^{\rm f}_C(\Omega,-\Omega,\Omega',-\Omega') $ are both smooth functions with respect $g_s^2(\omega_0-\Omega')g_s^2(\omega_0-\Omega)$, we have
\begin{equation}\label{eq:S2NG_fv2}
\begin{aligned}
&S_{n}^{(2){\rm NGf}}(\omega|\omega_0,\Delta \omega)\approx  \frac{2 (\Delta \omega)^2}{\pi^2} \Big[F_2   \tilde{G}^{\rm f}_C(\omega_0,-\omega_0,\omega_0,-\omega_0) \\
&+ \frac{F_0-F_2}{4}\tilde{G}^{\rm f}_B(\omega_0,-\omega_0,\omega_0,-\omega_0) \Big]~,
\end{aligned}
\end{equation}
where
\begin{equation}
\begin{aligned}
 \tilde{G}^{\rm f}_B(\omega_0,-\omega_0,\omega_0,-\omega_0)&= \frac{8 \ell_0}{L_x^3 L_y^3 \omega_0^3} \int^{\gamma_M/\omega_0}_{\Delta \omega/\omega_0} d \zeta  \frac{1-3\zeta^2}{(1+\zeta^2)^3}\\
&\approx \frac{8 \ell_0}{L_x^3 L_y^3 \omega_0^3} \int^{\infty}_{0} d \zeta  \frac{1-3\zeta^2}{(1+\zeta^2)^3}=0~,
\end{aligned}
\end{equation}
\begin{equation}
\begin{aligned}
 \tilde{G}^{\rm f}_C(\omega_0,-\omega_0,\omega_0,-\omega_0)&\approx  \frac{ \ell_0}{L_x^3 L_y^3 \omega_0^3} \\
 &\times \int^{\gamma_M/\omega_0}_{\Delta \omega/\omega_0} d \zeta  \frac{24}{(1+\zeta^2)(4+\zeta^2)}\\
&\approx \frac{ \ell_0}{L_x^3 L_y^3 \omega_0^3} \int^{\infty}_{0} d \zeta  \frac{24}{(1+\zeta^2)(4+\zeta^2)}\\
&=\frac{2 \pi \ell_0}{L_x^3 L_y^3 \omega_0^3}~. 
\end{aligned}
\end{equation}
Therefore, the non-Gaussian contribution to the second spectrum of the carrier density fluctuations induced by fast fluctuators reads
\begin{equation}\label{eq:S2NG_fvf}
\begin{aligned}
&S_{n}^{(2){\rm NGf}}(\omega|\omega_0,\Delta \omega)\approx  \frac{4 F_2 \ell_0}{ L_x^3 L_y^3} \frac{(\Delta \omega)^2}{\pi \omega_0^3} ~.
\end{aligned}
\end{equation}
It is white with respect to the variable $\omega$ like the corresponding Gaussian term, but it is proportional to $(\Delta \omega)^2$.
We conclude the calculation of the non-Gaussian part of the second spectrum with the contribution due to  slow fluctuators.
We notice that, because of the stationarity, we have $B(\gamma;t_1,t_2,t_3,t_4)=B(\gamma;t_1+t,t_2+t,t_3+t,t_4+t)$ and $C(\gamma;t_1,t_2,t_3,t_4)=C(\gamma;t_1+t,t_2+t,t_3+t,t_4+t)$, for $\forall t$.
According to the definition of the second spectrum, in 
$B(\gamma;t_1+\tau,t_2+\tau,t_3+\tau',t_4+\tau')$ and $C(\gamma;t_1+\tau,t_2+\tau,t_3+\tau',t_4+\tau')$ 
each time $t_i$, with $i=1,\ldots,4$, runs over the time interval $[-T_s/2, T_s/2]$, whereas $\tau$ and $\tau'$ run over the entire time interval $[-T_t/2,T_t/2]$, where  $T_t\to \infty$.
Thus, since the switching rates of the slow fluctuators are such that $1/T_t \ll \gamma \ll 1/T_s$, 
we can approximate  $B(\gamma;t_1+\tau,t_2+\tau,t_3+\tau',t_4+\tau')\approx B(\gamma;\tau,\tau,\tau',\tau')$ and $C(\gamma;t_1+\tau,t_2+\tau,t_3+\tau',t_4+\tau')\approx C(\gamma;\tau,\tau,\tau',\tau')$.
Under these conditions, we have
\begin{equation}\label{eq:S2NG_sv0}
\begin{aligned}
S_{n}^{(2){\rm NGs}}(\omega|\omega_0,\Delta \omega)&\approx \frac{8 \ell_0 g^4_s(\omega_0) }{L_x^3 L_y^3} 
\int_{\gamma_m}^{\Delta \omega}d\gamma\\
& \times  [1+(\ell_0/d) \ln(\gamma/\gamma_M)]^4\\
& \times \Bigg[\frac{2(F_2-F_0)}{\omega^4+4\gamma^2}+\frac{2F_2}{\omega^2+\gamma^2}\Bigg]~.
\end{aligned}
\end{equation}
By approximating $g^4_s(\omega_0)\approx 3[(\Delta \omega/(\pi \omega_0)]^4/8$, and neglecting corrections of order $(\ell_0/d)\ln[\gamma_M/(\Delta \omega)]\ll 1$, 
as done previously, we obtain
\begin{equation}\label{eq:S2NG_sv1}
\begin{aligned}
S_{n}^{(2){\rm NGs}}(\omega|\omega_0,\Delta \omega)&\approx \frac{3 \ell_0 }{L_x^3 L_y^3}  \frac{( \Delta \omega)^3}{\pi^4 \omega_0^3 \omega} 
\int_{\gamma_m/\omega}^{\Delta \omega/ \omega}d\zeta \\
& \Bigg[\frac{2(F_2-F_0)}{1+4\zeta^2}+
\frac{2F_2}{1+\zeta^2}\Bigg]\\
& \approx  \frac{3 \ell_0 }{L_x^3 L_y^3}  \frac{( \Delta \omega)^3}{\pi^4 \omega_0^3 \omega} [ 2 F_2 \arctan(\Delta \omega/\omega) \\
&+ (F_2-F_0)\arctan(2 \Delta \omega/\omega)]
~,
\end{aligned}
\end{equation}
moreover, by expanding for large $\Delta \omega/\omega$, we have
\begin{equation}\label{eq:S2NG_svf}
\begin{aligned}
S_{n}^{(2){\rm NGs}}(\omega|\omega_0,\Delta \omega)
&\approx   \frac{3 \ell_0 }{L_x^3 L_y^3}  \frac{( \Delta \omega)^2}{2 \pi^4 \omega_0^3}\bigg[\frac{\pi \Delta \omega }{\omega} (3 F_2-F_0)  \\
&+ (F_0-5 F_2)\bigg]~.
\end{aligned}
\end{equation}
The leading term of the non-Gaussian part of the second spectrum which involves slow fluctuators consists of a $1/\omega$ contribution and a $\omega$-independent term.  By comparing the white spectrum term of $S_{n}^{(2){\rm NGs}}(\omega|\omega_0,\Delta \omega)$ with $S_{n}^{(2){\rm NGf}}(\omega|\omega_0,\Delta \omega)$, it is straightforward to verify that the latter is dominant on the former. In fact, since  $5 F_2-F_0$ is of the same order of $F_0$, the $\omega$ independent term related to  slow fluctuators is smaller of  factor $3/(8 \pi^3)\approx 10^{-2}$ with respect to $S_{n}^{(2){\rm NGf}}(\omega|\omega_0,\Delta \omega)$.
By gathering up the contributions of the non-Gaussian part of the second spectrum of the carrier density fluctuations, we finally obtain
\begin{equation*}\label{eq:S2n_NG_final_SM}
\begin{aligned}
S_{n}^{(2){\rm NG}}(\omega|\omega_0,\Delta \omega)&\approx \frac{\ell_0}{L_x^3 L_y^3} \bigg[ 
\frac{4 (\Delta \omega)^2 F_2}{\pi \omega_0^3}+\frac{3 (\Delta \omega)^3}{2 \pi^3 \omega_0^3} \\
&\times \frac{3 F_2-F_0}{\omega} 
\bigg]~.
\end{aligned}
\end{equation*}
\vspace{0.1em}
%

%


\end{document}